# Isostructural phase transition in $Tb_2Ti_2O_7$ under pressure and temperature: Insights from synchrotron X-ray diffraction

Subrata Das,[1,*] Sanjoy Kr Mahatha,[2] Konstantin Glazyrin,[3]
R Ganesan,[1] Suja Elizabeth,[1] and Tirthankar Chakraborty[4,†]

[1]*Department of Physics, Indian Institute of Science, Bangalore 560012, India*
[2]*UGC-DAE Consortium for Scientific Research, Khandwa Road, Indore 452001, Madhya Pradesh, India*
[3]*Photon Sciences, Deutsches Elektronen-Synchrotron DESY, D-22607 Hamburg, Germany*
[4]*School of Physics and Materials Science, Thapar Institute of Engineering and Technology, Punjab 147004, India*

$Tb_2Ti_2O_7$, a pyrochlore system, has garnered significant interest due to its intriguing structural and physical properties and their dependence on external physical parameters. In this study, utilizing high-brilliance synchrotron X-ray diffraction, we conducted a comprehensive investigation of structural evolution of $Tb_2Ti_2O_7$ under external pressure and temperature. We have conclusively confirmed the occurrence of an isostructural phase transition beyond the pressure of 10 GPa. The transition exhibits a distinct signature in the variation of lattice parameters under pressure and leads to changes in mechanical properties. The underlying physics driving this transition can be understood in terms of localized rearrangement of atoms while retaining the overall cubic symmetry of the crystal. Notably, the observed transition remains almost independent of temperature. Our findings provide insights into the distinctive behaviour of the isostructural phase transition in $Tb_2Ti_2O_7$.

## INTRODUCTION

The pyrochlore family of materials has long been a subject of intense scientific interest due to its fascinating structural and physical properties [1] [2] [3] [4] [5]. These complex oxides, characterized by the $A_2B_2O_7$ stoichiometry, where A and B are cations, possess a pyrochlore crystal structure with a face-centered cubic lattice of B-site ions surrounded by A-site ions. This unique arrangement gives rise to a plethora of interesting phenomena, such as geometric frustration, exotic magnetic properties, and unusual electronic behaviour, making them a topic of immense interest in the scientific community [6] [7] [8] [9] [10] [11]. One particularly intriguing and novel feature of pyrochlores is the intricate relationship between phonons and the lattice, known as phonon-lattice coupling [8] [12] [13]. This phenomenon describes the mutual influence of lattice vibrations and the atomic structure of the crystal. In rare earth pyrochlores, the interplay between phonons and the underlying lattice leads to captivating effects, including phase transitions, lattice distortions, and novel electronic behaviours. This interplay becomes apparent in pyrochlore titanate systems, providing valuable insights into their complex behaviour [8] [14] [15] [16]. The coupling between phonon vibrations and the structural order implies that any changes in the phonon modes should be reflected in subtle modifications within the crystal structure. Indeed, these subtle changes in the crystal structure have been supported by the observed variations in lattice constants obtained from X-ray diffraction (XRD) measurements conducted at different temperatures [8]. The interplay between phonon dynamics and the underlying structural arrangement manifests itself in these tiny but significant alterations, providing concrete evidence of the intricate relationship between phonons and the crystal lattice within the pyrochlore titanate system.

Furthermore, the application of external pressure has been found to induce phase transitions in these systems [12] [17]. However, the nature and extent of these changes remain ambiguous and warrant further investigation. In a recent study by Ruff et al. [18], the lattice of $Tb_2Ti_2O_7$ showed anomalous expansion and broadening of Bragg peaks below 10 K, which was attributed to a cubic to tetragonal structural fluctuation associated with a correlated spin-liquid ground state. However, detailed investigations by Saha et al. [12] using Raman spectroscopy and XRD up to 10 K did not observe such features at low temperatures, raising questions about the material's behaviour and the reported ground state. Furthermore, both $Tb_2Ti_2O_7$ and $Gd_2Ti_2O_7$ have been subject to pressure-induced structural phase transitions which are reported to occur at pressures exceeding approximately 9 GPa [12] [17]. Interestingly, despite the occurrence of phase transitions, no discernible change in the structural symmetry or crystallographic space group was observed. Notably, there have been reports of pure cubic pyrochlores undergoing a transformation into a combination of tetragonal and cubic phases when subjected to increased pressure [19]. The reports on the structural changes induced by pressure in pyrochlores have therefore yielded inconclusive results. This intriguing aspect requires more elucidation and exploration which motivated us to undertake a comprehensive investigation into the pressure and temperature-induced phase transitions in these systems. We aim to shed light on the underlying mechanisms governing these transitions and contribute to a deeper understanding of the fascinating behaviour under variations of temperature and pressure.

To address these prospects, we conducted a compre-hensive series of experiments on $Tb_2Ti_2O_7$. Our investigation involved room-temperature Raman spectroscopy, laboratory-based XRD, and detailed synchrotron XRD under varying temperatures and pressure. We aimed to investigate any potential structural transitions and gain deeper insights into the material's behaviour and properties. Through rigorous analysis, we have unequivocally established the presence of a phase transition in the system occurring with variation in pressure. The transition has been identified as an isostructural phase transition, taking place at pressures beyond approximately 10 GPa. This involves significant changes in the material's structure, leading to drastic alterations in various physical properties, such as the bulk modulus, while the crystal's space group remains unchanged. Remarkably, we observed that this structural transition is essentially independent of temperature, whereas the bulk modulus experiences drastic changes within the temperature range of 130 to 160 K. In addition, we made efforts to shed light on the intricate interplay between phonon anharmonicity and its interaction with structural dynamics based on the observed results. Our study uncovers novel material properties and contributes to a better understanding of the underlying physics governing this system.

## EXPERIMENTAL DETAILS

Polycrystalline sample of $Tb_2Ti_2O_7$ was prepared using a solid-state synthesis route. Stoichiometric amounts of $Tb_2O_3$ and $TiO_2$ were thoroughly mixed and grounded using an agate mortar and pestle. After an initial sintering process at 1200 °C for 15 hours, the material was pelletized and underwent repeated sintering at 1400 °C for 15 hours each, with intermediate grinding steps in between. This process was carried out to achieve phase purity and better homogeneity of the sample. Room temperature XRD was performed using a state-of-the-art Rigaku system on the powdered sample and Raman measurements were carried out on a polished pallet of $Tb_2Ti_2O_7$ using a 532 nm laser line with 5 mW power in back-scattering geometry. The Raman spectroscopy was conducted within the range of 100 $cm^{-1}$ to 1000 $cm^{-1}$ using a Horiba setup. Temperature and pressure-dependent synchrotron XRD experiments were carried out at the P02.2 beamline of PETRA III, DESY, Germany with an X-ray beam of wavelength 0.483 Å (25.8 keV). During the pressure-dependent synchrotron XRD measurements, the pressure was varied from ambient to 34 GPa using a membrane-type diamond anvil cell, with helium as the pressure media and Rb crystals were employed to calibrate the pressure. The temperature was varied from 92 K to 250 K using a liquid helium cryostat.

TABLE I: Structural parameters of $Tb_2Ti_2O_7$ obtained from Rietveld refinement of XRD pattern at ambient pressure and temperature.

| Space group | $Fd\bar{3}m$ |
|---|---|
| Lattice parameters | $a$=$b$=$c$=5.518 Å |
| Tb – O1 | 2.58(3) Å |
| Tb – O2 | 2.19(2) Å |
| Ti – O2 | 1.91(0) Å |
| Tb – O1 – Tb | 109.47(1)° |
| Tb – O2 – Tb | 87.70(6)° |

## RESULTS AND DISCUSSIONS

### Structural Characterization

The room-temperature XRD pattern obtained for $Tb_2Ti_2O_7$ is presented in Figure 1 (a). The observed data is refined using the Rietveld method [20] with the cubic space group $Fd\bar{3}$m, employing FullProf software [21]. The calculated pattern exhibits excellent agreement with the experimental data, resulting in a global $\chi^2$ value of 2.56. The absence of any impurity peaks and a good $\chi^2$ value confirm the formation of the desired phase without any impurity. The schematic of the crystal structure is deduced using VESTA software, based on the refined lattice parameters extracted from the XRD data, and is shown in the inset of the figure 1 (a). The structural parameters obtained from the fitting process are summarized in Table I. In the crystal structure, $Tb^{3+}$ and $Ti^{4+}$ occupy the 16d and 16c Wyckoff positions respectively. Additionally, two distinct types of oxygen atoms are identified and denoted as $O$ and $O'$, which occupy the 48f and 8b Wyckoff positions, respectively. According to the group theory analysis of the space group $Fd\bar{3}m$, the vibrational modes of $Tb_2Ti_2O_7$ are predicted as $A_{1g} + E_g + 2F_{1g} + 4F_{2g} + 3A_{2u} + 3E_u + 7F_{1u} + F_{2u}$. Among these modes, $A_{1g}$, $E_g$, and $F_{2g}$ are Raman active, while $F_{1u}$ and $F_{2u}$ are infrared active [17] [12]. Figure 1(b) shows the Raman spectra for $Tb_2Ti_2O_7$ at room temperature and ambient pressure measured in the range of 100 to 1000 $cm^{-1}$. Deconvolution of the recorded spectra with Lorentzian obtains seven Raman modes labelled as $M1$ to $M7$. Details of these modes, including their respective wave numbers, intensities, and assignments, are provided in Table II, which closely matches with the previous reports [17] [22].

### Pressure and temperature dependence of structure

To investigate and gain insights into the possibility of any phase transition with pressure, we performed pressure-dependent synchrotron XRD measurements at various temperatures. In Figure 2 (a), we present the 2D diffractogram acquired via synchrotron XRD at two specific pressures: 5 GPa and 28 GPa. Notably, the blue arrows pinpoint the presence of the most intense peak, identified as (222), while the red arrows draw attention to the bright spots caused by the use of helium as a pressure-stabilizing medium. The XRD patterns are derived from the 2D diffractograms acquired under varying temperature and pressure conditions. It is worth noting that this bright spot becomes more pronounced and gradually expands into a circular shape as the pressure increases, which is clearly reflected in the XRD patterns shown in Figure 2 (b), within the $2\theta$ range of 13° to 15°, as indicated by the blue stars. As a representative example, Figure 2(b) displays the evolution of Bragg peaks as a function of applied pressure at a temperature of 94 K. The pressure dependence of the most intense peak (222) is shown in Figure 3. The peak is observed to shift towards higher $2\theta$ values as pressure increases, indicating a reduction in the lattice cell dimensions. To quantitatively analyze the changes, the peaks were fitted using the Voigt function. The full width at half maximum (FWHM) of the peaks were determined and found to increase with increasing the applied pressure. This observation suggests that the sample experiences increasing strain as pressure is applied, resulting in lattice distortion.

TABLE II: Raman modes of $Tb_2Ti_2O_7$ and its positions obtained by deconvoluting the spectrum.

| Modes | position ($cm^{-1}$) | Assignment |
|---|---|---|
| $M_1$ | 221.02 | $F_{2g}$ |
| $M_2$ | 307.81 | $F_{2g}$ |
| $M_3$ | 355.63 | $E_g$ |
| $M_4$ | 521.99 | $A_{1g}$ |
| $M_5$ | 543.78 | $F_{2g}$ |
| $M_6$ | 686.38 | $F_{2g}$ |
| $M_7$ | 714.39 | *Second order Raman mode* |

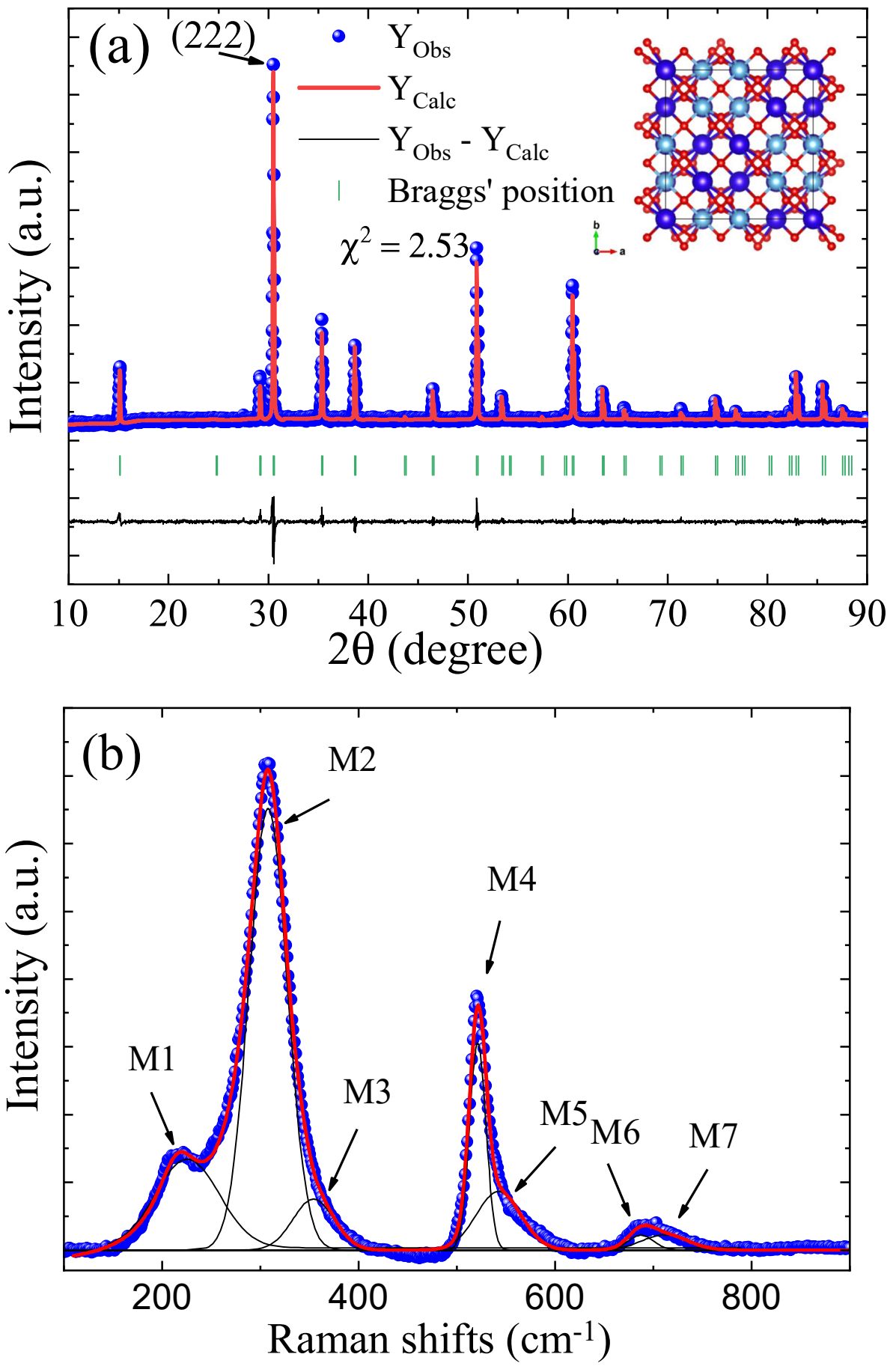

FIG. 1: (a) Room temperature powder XRD pattern ($\lambda$ = 1.54 Å)of $Tb_2Ti_2O_7$ with Rietveld refinement utilizing the space group $Fd\bar{3}m$. Inset shows the crystal structure deduced from refined data using VESTA software. (b) Room temperature Raman data of $Tb_2Ti_2O_7$. Deconvoulated peaks are assigned to different modes of vibration.

Notably, upon careful examination of the XRD patterns, it is evident that no peaks merge, split, or appear with increasing pressure at any temperature. This observation strongly suggests that there is no change in the symmetry of the system under pressure, which is consistent with previous reports [8] [12]. Given this scenario, initially, it raises questions about the occurrence of a drastic structural transition beyond the usual gradual changes in the lattice structure under pressure. However, it has been reported that above 9 GPa, the disappearance of phonon modes $F_{2g}$ upon decompression indicates some abrupt changes in the lattice beyond this pressure, suggesting the possibility of a structural transition [18]. To get deeper insight into this, lattice parameters at all pressures and temperatures are extracted using Rietveld refinement of the XRD patterns and analysed in detail. Figure 4 illustrates the variation of lattice parameters with increasing pressure up to approximately 34 GPa at different temperatures. It is evident that both the lattice constant and volume exhibit a continuous decrease as pressure is applied. However, a noteworthy anomaly is observed at around the onset pressure of 10 GPa. Remarkably, this anomaly is observed at all temperatures and appears to be independent of temperature. Such a consistent behaviour suggests the occurrence of a significant structural transition at pressure 10 GPa. Furthermore, the pressure-dependent changes in volume can be fitted using the third-order Brich-Murnaghan equation of state [23] given by Equation 1 as follows:

$$P = \frac{3}{2}B_0\left[\left(\frac{V}{V_0}\right)^{-7/3} - \left(\frac{V}{V_0}\right)^{-5/3}\right] \left\{1 + \frac{3}{4}(B_0' - 4)\left[\left(\frac{V}{V_0}\right)^{-2/3} - 1\right]\right\} \tag{1}$$

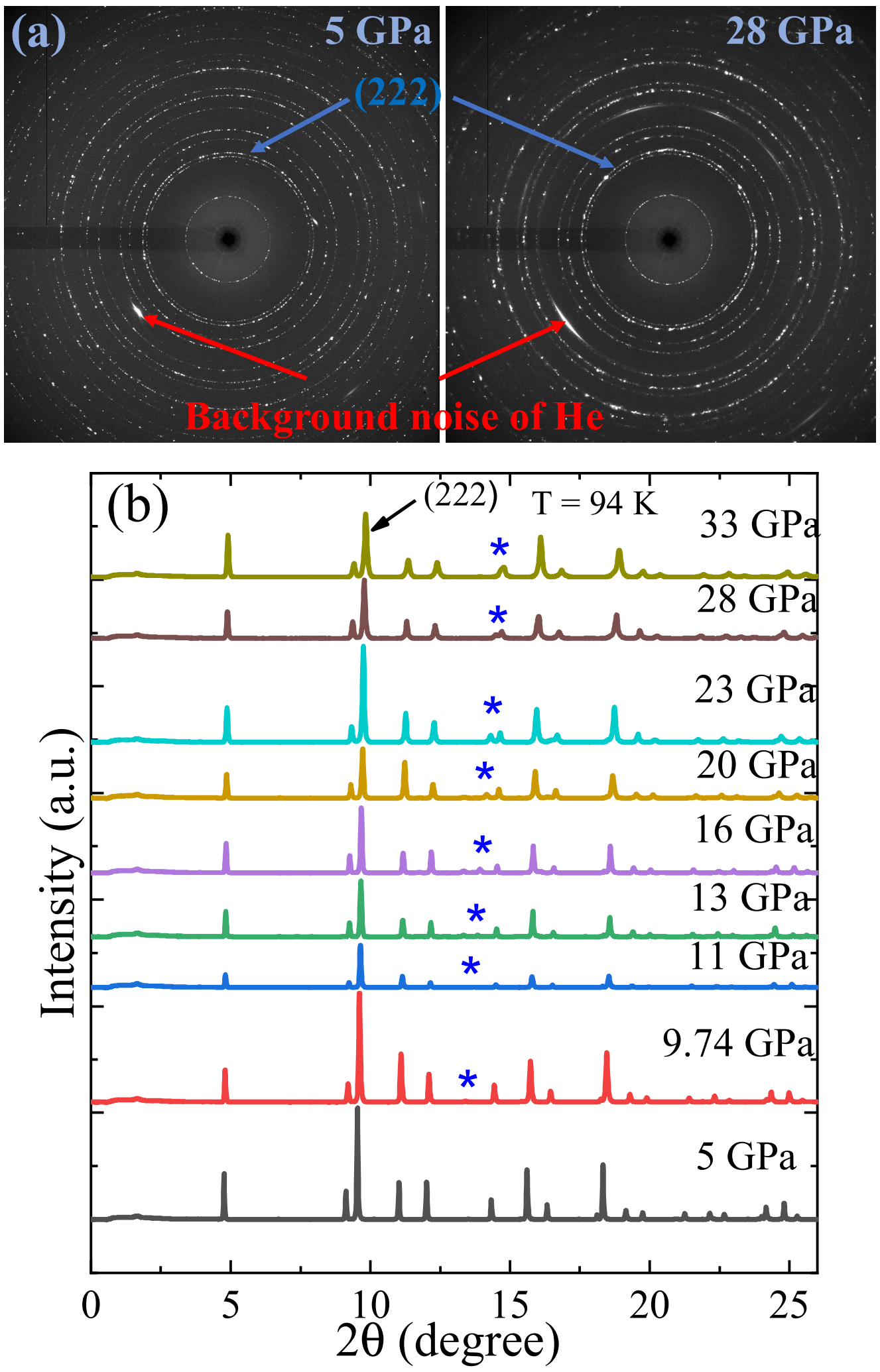

FIG. 2: (a) The 2D diffractogram obtained at pressures of 5 GPa and 28 GPa exhibits a noticeable (222) peak, with concurrent background noise arising from the employment of helium as the pressure medium. (b) Synchrotron XRD patterns at 94 K with pressure up to 34 GPa.

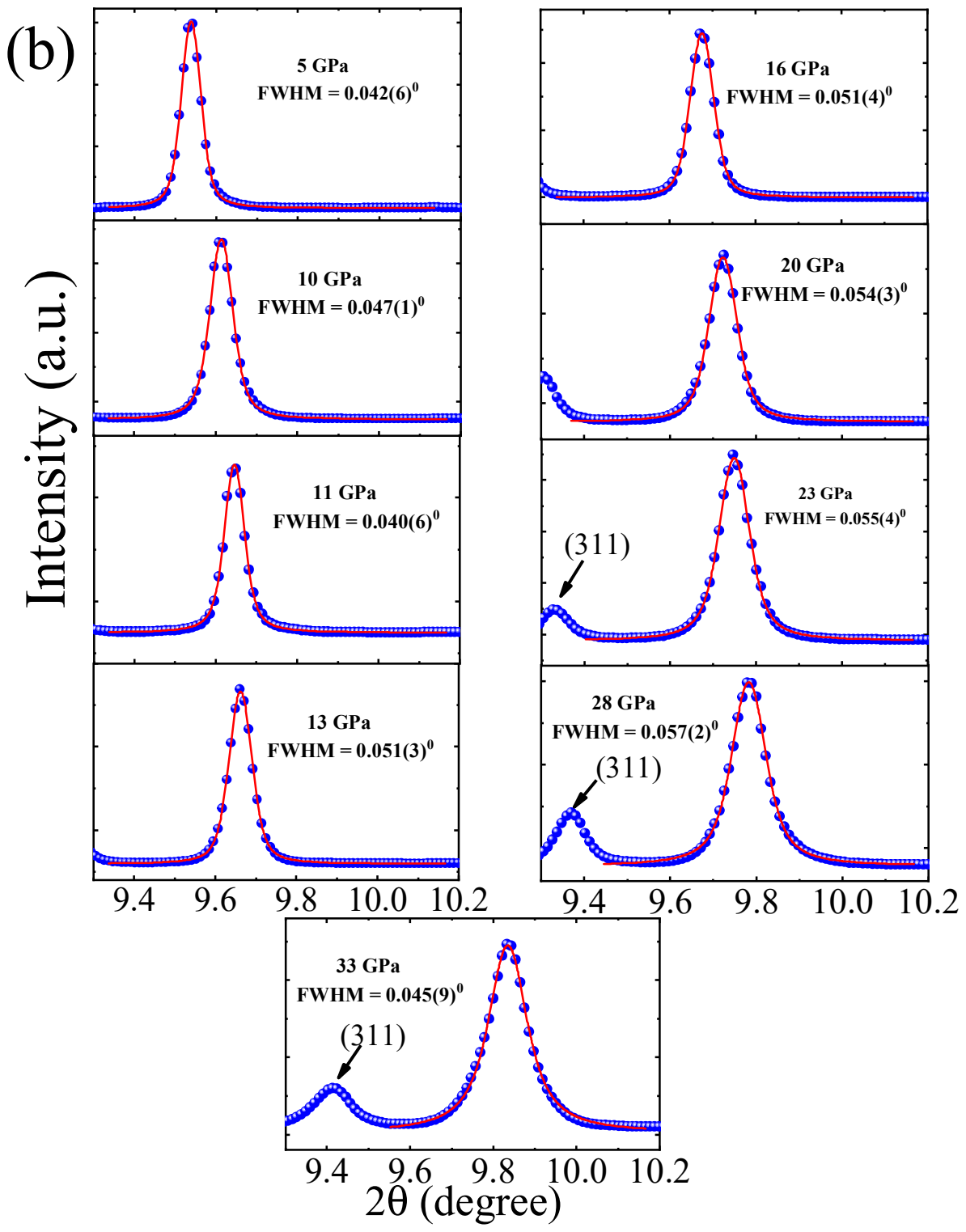

FIG. 3: Evolution of the (222) peak with pressure along with fitting. Broadening of the peak upon increasing pressure is noticeable.

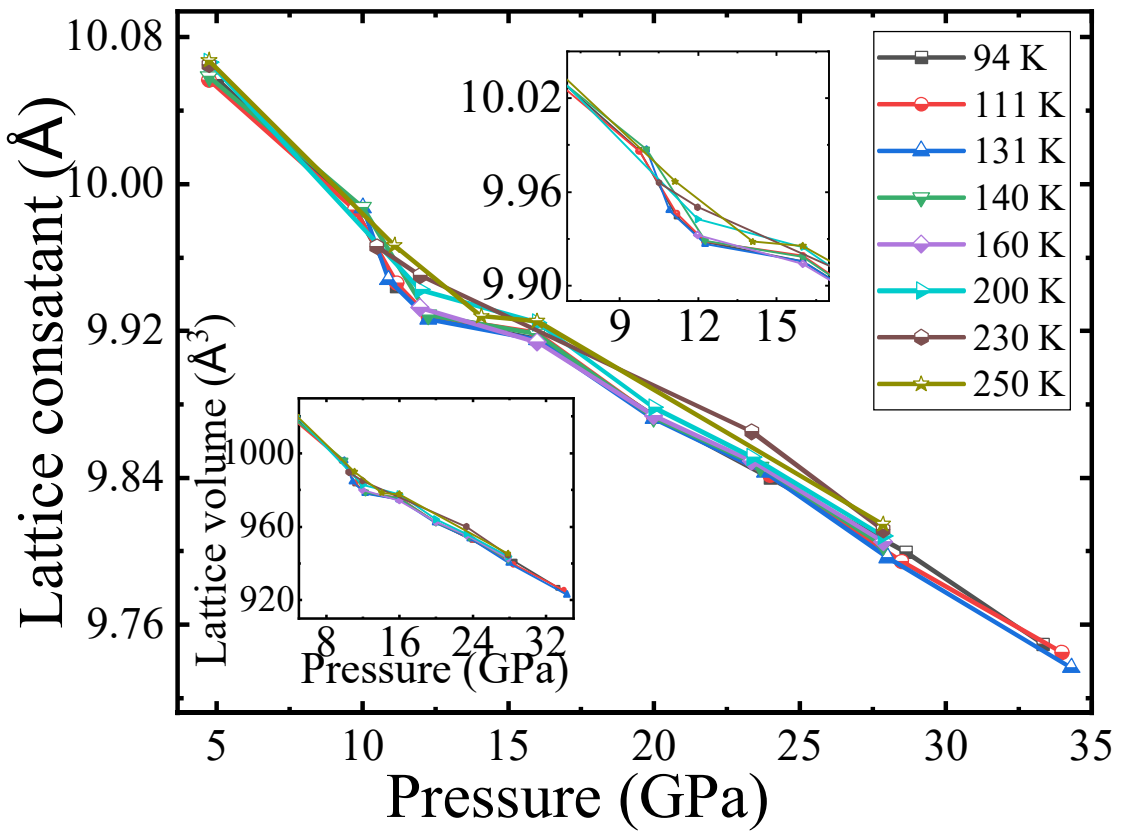

FIG. 4: Variation of the lattice constant with pressure at different temperatures, top inset is showing enlarged view near the anomaly. The bottom inset shows the variation of volume with pressure at different temperatures.

where, $B_0$ and $B_0^{'}$ are the bulk modulus and 1st derivative of the bulk modulus with respect to pressure at ambient pressure, respectively. At 94 K, this fitting yields the bulk modulus, $B = 227.4\ GPa$ below 10 GPa, while above the anomaly at around 15 GPa, these values change to $B = 293.9\ GPa$. This indicates 29.24% increase in the bulk modulus across the anomaly. The observed anomaly, in conjunction with the significant variation in the bulk modulus, provides strong evidence of a drastic change in the structure around the anomaly. It is important to note that in certain cases [17] [12], a methanol-ethanol mixture is employed as a hydrostatic pressure medium which becomes non-hydrostatic around 11 GPa, leading to the possibility of anomalies in the experimental results. In our study, however, we have ruled out this possibility by using helium as a hydrostatic pressure stabilizing medium. This ensures that our experimental conditions remain hydrostatic throughout the pressure range, enhancing the reliability and accuracy of our findings. The structural transition observed in the system leaves the overall cubic symmetry of the crystal intact. This phenomenon is known as an isostructural phase transition. The mechanism behind this isostructural phase transition can be explained in the context of the lattice dynamics and the behaviour of the constituent atoms under pressure. As pressure is applied, the interatomic distances and bond angles in the crystal lattice are altered, leading to changes in the atomic positions and local coordination environment. In pyrochlore compounds like $Tb_2Ti_2O_7$, the crystal structure consists of a three-dimensional network of corner-sharing tetrahedra, where the A-site cations (in this case, $Tb^{3+}$) occupy the centre of each tetrahedron, and the B-site cations ($Ti^{4+}$) are located at the corners[17] [24, 25]. The oxygen ions form the corners of the tetrahedra and are shared between neighbouring tetrahedra. As pressure is applied to the material, the interatomic distances and bond angles are altered. In the case of $Tb_2Ti_2O_7$, the rare-earth cations ($Tb^{3+}$) have relatively large ionic radii, and the oxygen ions are more compressible than the metal cations. This should make the rare-earth cations and oxygen ions more responsive to changes in pressure compared to the titanium cations. Therefore, the transition is likely driven by the response of the rare-earth cations ($Tb^{3+}$) and the oxygen ions. It is important to note that attempting to extract the values of fractional coordinates of atoms, bond angles, and bond lengths from Rietveld refinement of XRD patterns at different temperatures and pressures could be misleading. Refining oxygen positions may not yield accurate results, as oxygen contributes relatively less to the scattered X-rays. However, it is logical to infer that under pressure, the rare-earth cations tend to shift slightly, leading to local distortions in the oxygen coordination polyhedra around them. While there are substantial distortions at the atomic scale, the local distortions are not sufficient to break the overall symmetry of the crystal lattice. Instead, they result in a dynamic interplay of local structural changes, where some bonds are compressed while others are elongated, effectively maintaining the cubic symmetry at a macroscopic level. Systematic studies with electron diffraction can be important in this context to provide further confirmation of this mechanism. In $Tb_2Ti_2O_7$, the isostructural transition occurs at approximately 10 GPa, coinciding with the disappearance of specific phonon modes, particularly the $F_{2g}$ modes, observed in previous studies [17]. This indicates coupling between phonons and structural dynamics which is crucial in this context. The changes in atomic positions and local coordination induced by pressure affect the phonon frequencies and anharmonicity, and in turn, the anharmonic phonons contribute to the atomic rearrangements. This mutual influence between phonon anharmonicity and structural dynamics leads to the observed isostructural transition, where localized rearrangements occur while preserving the overall cubic symmetry of the system.

Furthermore, we conducted an investigation into the temperature dependence of the structural and mechanical properties. Figure 5 illustrates the variation of the bulk modulus with temperature. Notably, the bulk modulus undergoes a significant change within the temperature window of 130 to 160 K, indicating a distinct temperature-dependent behaviour at this range. However, above 160 K, the bulk modulus shows only a very slight temperature dependence, remaining essentially constant. Similarly, below 130 K, the bulk modulus remains almost unchanged with temperature. Whereas, $B_0^{'}$ shows reverse trends with a significant change in the similar temperature range. This is attributed to the rearrangement of the $TiO_6$ octahedra around this temperature. The value of $B_0^{'}$ is less at lower temperatures and increases after 130 K indicating the increase of $TiO_6$ distortion [19]. The expansion of the lattice with temperature can be well-described by the equation:

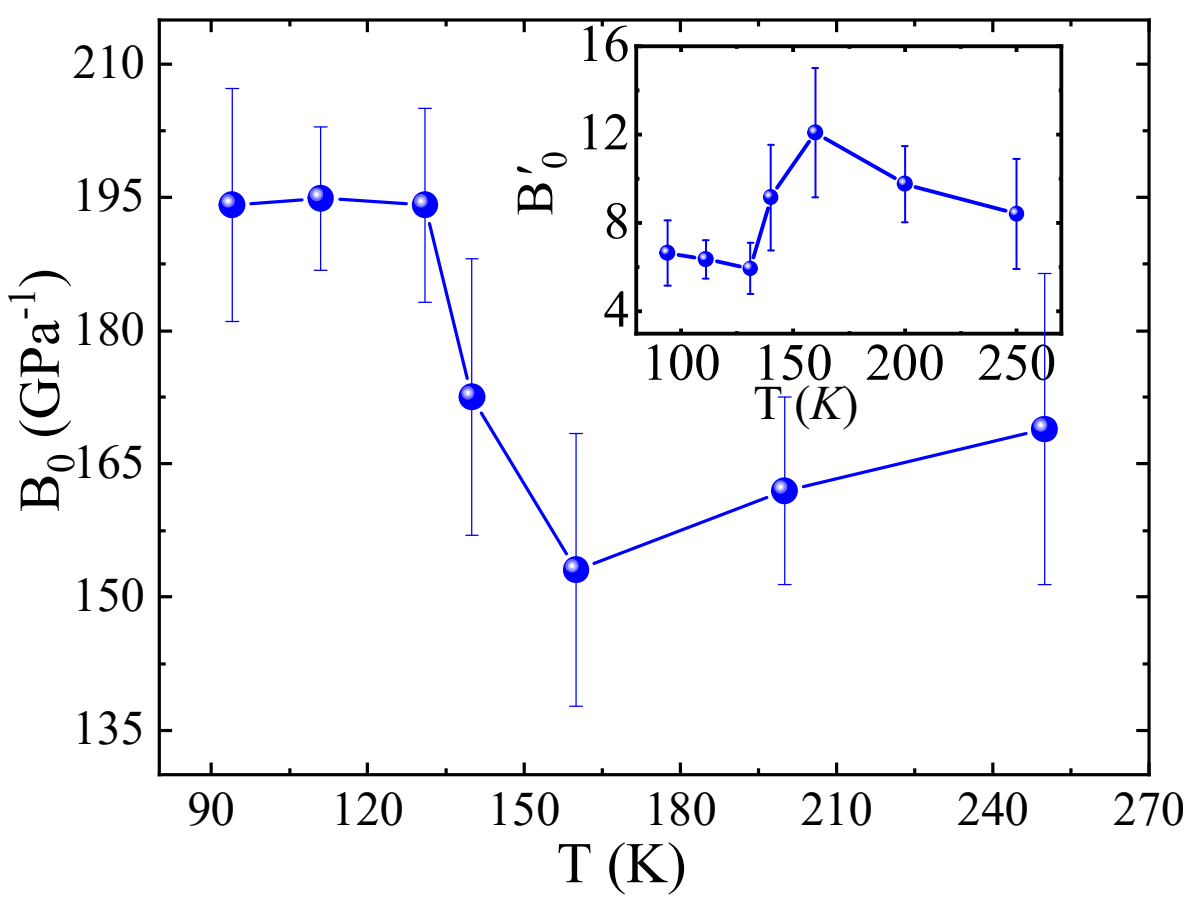

FIG. 5: Variation of bulk modulus at ambient pressure ($B_0$) with temperature, inset shows variation of 1st derivative of bulk modulus ($B_0^{'}$) with temperature.

$$a(T) = a(T_0)\left[\frac{b\exp\left(\frac{c}{T-T_0}\right)}{T\left(\exp\left(\frac{c}{T-T_0}\right)-1\right)^2}\right] \quad (2)$$

Here, $a(T_0)$ is the lattice constant at 0 K, $a$ and $b$ are the fitting parameters. Figure 6 presents the fitting of the experimentally obtained lattice constants with this equation for three representative pressures of 12 GPa, 23 GPa, and 28 GPa, respectively. The fitting parameters are listed in Table III. The expansion coefficient $a_v$ can be calculated from the slope as $a_v = \frac{3}{a_0}\left(\frac{da}{dT}\right)$. The variation of the lattice parameters with temperature is similar to that reported by Saha et al. [12]. Figure 6 (d) shows the variation of the expansion coefficient with pressure. The expansion coefficient exhibits an increasing trend with pressure, followed by a sharp decrease beyond 11 GPa, before showing a slight increase again. This pronounced decrease is associated with the isostructural phase transition, consistent with the observations in Figure 4. This comprehensive analysis allows us to quantify the lattice expansion of the material with temperature under dif-ferent pressure conditions. Furthermore, a detailed and comprehensive pressure and temperature-dependent Raman study may draw a clear picture of the anomalous behaviour of phonons with the help of the Grüneisen parameter.

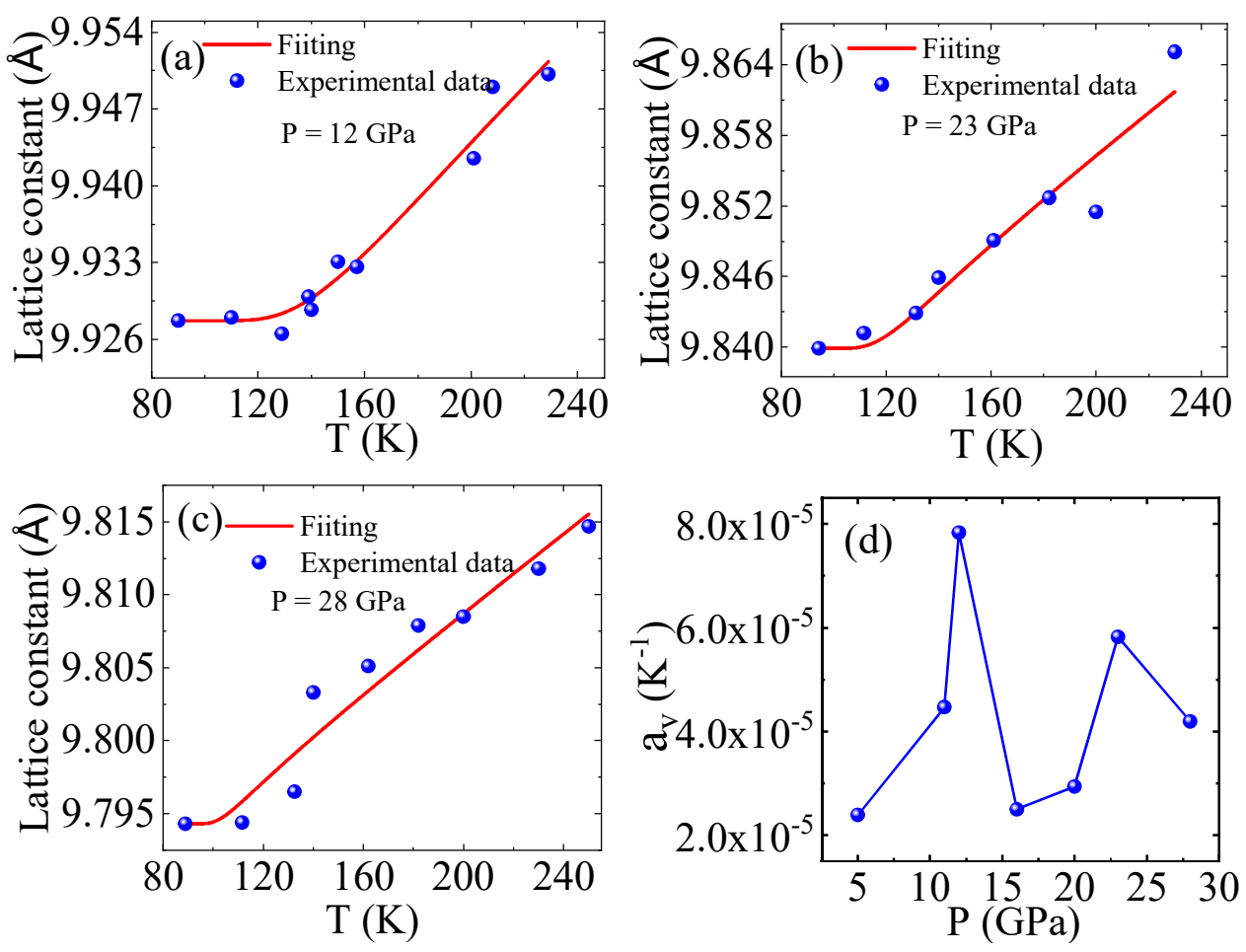

FIG. 6: Variation of the lattice constant with temperature at different pressures: (a) 12 GPa, (b) 23 GPa and (c) 28 GPa. Solid red curves represent fittings with Equation 2. (d) Variation of the expansion coefficient ($a_v$) with pressure.

TABLE III: Structural parameters of $Tb_2Ti_2O_7$ obtained from Rietveld refinement of XRD pattern.

| Pressure | b | a |
|---|---|---|
| Ambient | 9.45 K | 648.5 K [17] |
| 12 GPa | 1.12778 K | 241 K |
| 23 GPa | 0.2238 K | 112.74 K |
| 28 GPa | 0.0609 K | 66.72 K |

## CONCLUSION

In conclusion, we prepared $Tb_2Ti_2O_7$ and carried out a comprehensive investigation under varying temperatures and pressure. The results from synchrotron XRD confirm the occurrence of an isostructural phase transition beyond approximately 10 GPa. This transition is evident from the anomalies observed in the variation of lattice parameters and changes in the mechanical properties of the system with pressure. It is likely to be driven by the response of the rare-earth cations ($Tb^{3+}$) and the oxygen ions to pressure-induced changes in the lattice and can be explained in terms of localized rearrangements these atoms while preserving the overall cubic symmetry of the crystal. The interplay between phonon anharmonicity and its interaction with structural dynamics plays a crucial role in the observed phenomena. Moreover, it is noteworthy that the observed transition is essentially temperature-independent. Overall, our study contributes to the understanding of the complex behaviour of $Tb_2Ti_2O_7$ and the underlying physics driving its phase transitions.

## ACKNOWLEDGEMENT

Synchrotron diffraction studies were performed at the light source PETRA III of DESY, a member of the Helmholtz Association. Financial support from the Department of Science and Technology, Government of India provided within the framework of the India@DESY collaboration is gratefully acknowledged.

* Electronic address: subratadas@iisc.ac.in
† Electronic address: tirthankar@thapar.edu